\documentclass[conference]{IEEEtran}
\ifCLASSINFOpdf
   \usepackage[pdftex]{graphicx}
\else
\fi
\usepackage[cmex10]{amsmath}
\begin{document}
%
\title{Novel velocity model to improve indoor localization using inertial navigation with sensors\\ on a smart phone}


\author{\IEEEauthorblockN{Hettiarachchige Don Rasika Lakmal, Jagath Samarabandu} 
\IEEEauthorblockA{Department of Electrical and Computer Engineering\\
University of Western Ontario, Canada}}


%


\maketitle

\begin{abstract}
We present a generalized velocity model to improve localization when using an Inertial Navigation System (INS). This algorithm was applied to correct the velocity of a smart 
phone based indoor INS system to increase the accuracy by counteracting the accumulation of large drift caused by sensor reading errors. We investigated the accuracy of the 
algorithm with three different velocity models which were derived from the actual velocity measured at the hip of walking person. Our results show that the proposed method 
with Gaussian velocity model achieves competitive accuracy with a 50\% less variance over Step and Heading approach proving the accuracy and robustness of proposed method. We also 
investigated the frequency of applying corrections and found that a minimum of 5\% corrections per step is sufficient for improved accuracy. The proposed method is applicable in indoor 
localization and tracking applications based on smart phone where traditional approaches such as GNSS suffers from many issues.
\end{abstract}


%
\IEEEpeerreviewmaketitle

\section{Introduction}
Localization and tracking of entities has always been an essential part in many application domains. With the emergence of Internet Of Things (IOT), knowing the location of each 'thing' will be a key requirement towards providing the best benefit \cite{perera2014context}. Global Navigation Satellite Systems (GNSS) which covers the whole world through satellites have matured for outdoor use. But for indoor use, GNSS suffers from serious issues due to high signal attenuation or complex signal conditions \cite{do2014smartphone}. To address the important need of localization and tracking indoors, many approaches have been researched and implemented as summarized below \cite{harle2013survey}.  \\

\begin{itemize}
\item{Specialized hardware and infrastructure}\\
These methods install special infrastructure or utilize existing infrastructure inside buildings for the purpose of localization. Tracking and localization is mostly achieved using triangulation from known locations, based on the Time of Arrival(TOA) or Angle of Arrival(AOA) techniques. Examples include systems based on RFID tags, Bluetooth or Wifi equipped devices.
\item{Signal Fingerprinting}\\
These methods map and store the signal properties (e.g. Received signal strength) of one or more types (e.g. RF, magnetic) of signals within a given area and then use that signal map to localize in realtime by comparing the current signal property. Many advancements have been researched to improve the accuracy of comparison and to simplify the process of map building. This method is currently gaining popularity with commercial applications.
\item{Dead reckoning}\\
These methods measure the displacement using available sensors when started from a known position. This is common in many applications with the displacement being calculated by acceleration signals from the inertial sensors. Dead reckoning for firefighters with foot mounted sensors is a common application in this category.
\end{itemize}

Dead reckoning attempts for localization and tracking of pedestrians can be grouped in to a) Inertial Navigation Systems(INS) and b) Step and Heading(SHS) Systems \cite{harle2013survey}.
INS systems attempt to continuosly track the location of a sensor with reference to an initial point in 3D. Tracking of location of a user by means of dead reckoning based on sensor readings from a wearable device is a common application in this category. Acceleration and gyroscope sensors which measure acceleration and the rotation velocity respectively are commonly used where the second integral of acceleration yeilds the distance and first integral of rotation velocity yeilds the direction. While in theory, this method should generate a 3D trajectory of the sensor unit, in practice, an increasingly large error (drift) will arise due to the errors in the sensor readings. In order to minimize this drift, corrections are applied during the localization process with some external measurements. A commonly known technique called Zero Velocity Updates (ZUPTs) clamps velocity to zero when a stationary phase is detected in a foot mounted sensor application \cite{foxlin2005navshoe}. Application of corrections while tracking the error covariance in a Kalman filter is commonly accepted rather than directly setting the velocity to zero.\\
Step and Heading Systems (SHS) attempt to identify the steps and predict the length of the step by processing a signal from a wearable sensor of a walking user and then orient that step length 
along the direction calculated by a magnetometer or gyroscope. Main components of this method are step detection and step length calculation for which many algorithms have 
been suggested in literature \cite{seco2009comparison}. \\
With the popularity of the smart phones which incorporates a variety of advanced sensors, indoor localization and tracking based on smartphone is gaining popularity in 
academia and industry. Although fingerprinting methods based on wifi and cellular signals are currently popular with the smart phone applications, a trend can be observed in 
usage of inertial sensors available in smart phones for dead reckoning. These systems can perform well in applications by initializing from a known location. Hybrid systems which employ combinations of INS, SHS and fingerprinting methods also has been proposed to overcome inherent disadvantages of each approach \cite{lukianto2010pedestrian}\cite{chen2014intelligent}. The SHS seem to be the de facto approach in smart phone based indoor applications \cite{kang2015smartpdr}. This fact is observed in the recent survey by Subbu et al.   \cite{subbu2014analysis} and applications by Do-Xuan et al. \cite{do2014smartphone}, Kang et al. \cite{kang2015smartpdr} and Park et al. \cite{park2015robust}.\\

In INS dead reckoning, Zero velocity updates are used to correct the drift with foot mounted sensor applications when the foot is in ``Stance'' phase (the interval between heel strike and toe-off where foot is in contact with the ground). In a smart phone application, the sensors need to be placed around hip area.Since a hip mounted smart phone does not come to a complete stop in the stance phase, ZUPT method yeilds unacceptable levels of error when applied in this scenario. \\
In this paper, we propose a novel approach for applying corrections to INS based localization algorithm using a velocity model. In this method, measurements from a velocity model derived from actual measurements are employed as external corrections. We investigate three different velocity models derived from Gaussian, Sinusoidal and sawtooth functions for suitability in the proposed method. The corrections are applied in a complementary Kalman filter since it allows providing a measure of uncertainty of the correction along with the correction itself. In this way, both the measured acceleration components and the model will compliment each other. Our experiments show that the proposed approach improves upon existing SHS methods for localization using a smart phone. \\

\section{Methodology}
In this experiment, walking trials with a hip mounted smart phone were performed and accuracy of proposed INS method and current SHS method was compared. First part describes the main tracking application and the second part describes the generation of three different velocity models to approximate the actual curve. In this experiment, for the purpose of comparison, straight walks along a flat surface were performed.\\

\subsection{Strapdown INS}
The smart phone was placed firmly with vertical orientation in a belt pocket of the person performing the walking trials. While walking, sensor readings were accessed and transmitted to a server using the data connectivity in the phone and the analysis was performed.  
Android operating system based smart phone with in built tri-axel accelerometer and gyroscope was used in this experiment and the access to the sensor readings was achived thorugh the Operating System's Application Programming Interface(API). Android API allows the use of a gravity sensor which provides the gravitational acceleration components, a linear acceleration sensor which provides the acceleration without gravity component and a gyroscope sensor which provides the angular velocity and access. These sensor's data can be read at a maximum frequency of 100Hz \cite{android}.\\
The axes on which the sensor readings are based with reference to the phone body, which we call the phone frame, is depicted in the figure 1. 

\begin{figure}[!t]
\centering
\includegraphics[width=1.0in]{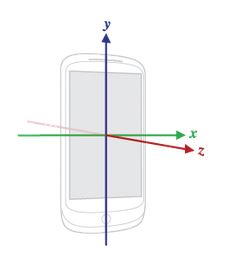}
\caption{Cordinate system on phone frame. \cite{android}}
\end{figure}

Due to the fact that orientation of the phone can be offset relative to the ground or navigation frame, and this offset may change while walking, a transformation is needed to convert the phone frame readings to the navigation frame by using Euler angles of the phone which are called Yaw, Pitch and Roll. Since the phone was fixed in vertical position with freedom to move around X and Z axes, the roll angle was assumed negligeble and set to zero. The pitch and yaw angles were calculated from the gravity sensor by extending the approach suggested by Do-Xuan et al. \cite{do2014smartphone} for positive and negative angles of yaw and pitch.\\
Once the navigation frame acceleration components were calculated, the acceleration along the walking direction was integrated once to get the velocity, and again velocity is integrated to get the distance. In both cases, the trapezoidal numerical integration was used.\\
This process, which is summarized in steps below is the standard strapdown INS technique.
\begin{itemize}
\item{Read sensor data from the device}
\item{Calculate the orientation matrix}
\item{Transform the sensor data in to navigation frame}
\item{Integrate twice to get the displacement}
\end{itemize}

Due to the bias and random errors present in the sensor readings, the calculated distance deviates from the actual distance exponentially within few seconds. In order to correct that, zero velocity updates have been employed in the foot mounted inertial sensor scenario when the foot is in the stance phase. Rather than directly resetting the velocity to zero, the error is maintained as the state in Kalman filter and correction is applied as an measurement to the filter. This is the known as complementary Kalman filter application. This method allows to \cite{woodmanthesis},
\begin{itemize}
\item{Include an uncertainity of the external measurement in to the calculation}
\item{Correct the position and other related predictions using the relation their with velocity}
\end{itemize}
Instead of the zero velocity, application of a velocity correction measurement from an actual model was proposed and experimented here. 
The Kalman filter maintains the probability distribution of the errors in the strapdown INS in the state as below where \(\delta p\), \(\delta v\),\(\delta \theta\) and \(\delta \omega\) denotes position, velocity, heading and angular velocity errors respectively.

\[ \delta x = \left| \begin{array}{cccc}
\delta p &  \delta v & \delta \theta & \delta \omega \\
 \end{array} \right|'\] 

In order to simplify the filter implementation, only velocity and distance along the walking direction was considered in the error state. \\
Then the state and the error covariance is propagated with the INS measurements using the standard kalman filter equations with the following state trasition matrix which relates the velocity errors to the distance errors. (\(\delta t\) is the difference between two samples, which is 0.01 seconds in our experiments) \\
\[ F = \left| \begin{array}{cccc}
1 & 0 & 0 & 0 \\
\delta t & 1 & 0 & 0 \\
0 & 0 & 1 & 0 \\
0 & 0 & 1 & \delta t \end{array} \right|\] 
When an external measurement is available, the error between the current estimate and the external measurement is calculated. This is applied as a measurement to the Kalman filter which updates the error state and the error covariance using standard kalman filter equations with the following measurement matrix. \\
\[ H = \left| \begin{array}{cccc}
0 & 1 & 0 & 0 \\
0 & 0 & 1 & 0 \end{array} \right|\] 
Once the correction phase is completed, error values are transfered to the underlying INS which corrects the position and velocity estimate using the errors calculated by kalman filter.

\subsection{Actual velocity model}
The actual hip velocity was measured using Hagisonic indoor localization system \cite{hagisonic}. The system outputs the position measurements signal which was differentiated using central approximation to get the velocity. After recognizing the pattern for each step and with the combination of acceleration integrated velocity, model parameters were selected to map the velocity.\\ 
Three models derived from Gaussian, Sinusoidal and Trapezoidal functions were evaluated in our experiment.\\
\subsubsection{Gaussian Model}
This model is shown in equation (1) where $T$ is the step period, $A$ is the velocity shift, $K$ is a constant and $a$ is the mean fraction constant which changes the peak velocity point. The variance \(\sigma\) was defined as a fraction of the step period in order to maintain the shape of the velocity curve across steps. So the value of variance was defined as \(\sigma = b*T \) where $b$ is the step fraction constant. 
\begin{equation}
V(t) = A +\frac{ K}{T}\exp^{\frac{-(t-aT)^2}{2 \sigma^2}} 
\end{equation}
\subsubsection{Sinusoidal Model}
This model is shown in equation (2), where $T$ is the step period, $a$ is the amplitude and $K$ is the shift scale factor which are constants.
\begin{equation}
V(t) = K*T + a*sin(\frac{2 \pi t}{T}) 
\end{equation}
\subsubsection{Sawtooth Model}
Sawtooth model is shown in equation (3), where $T$ is the step period, $a$ is the amplitude, $K$ is the shift scale factor and $b$ is the width which defines the peak point in the signal.
\begin{equation}
V(t) = K*T + a*sawtooth(2 \pi t, b) 
\end{equation}
Applying the velocity model for correction in strapdown INS requires detecting each step. From the many algorithms available to detect steps from the acceleration signal, we used the Finite State Machine (FSM) method in this experiment \cite{alzantot2012uptime}. This algorithm has been proven to be more accurate when used with a smart phone \cite{alzantot2012uptime}. Figure 2 shows this algorithm, where 4 different threshold values are used to switch between different phases of the step.
 
\begin{figure}[!t]
\centering
\includegraphics[width=2.8in]{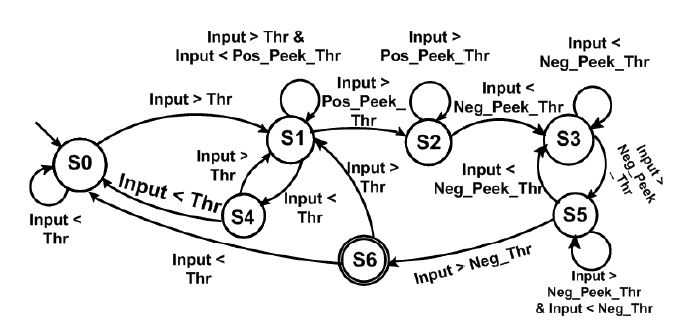}
\caption{FSM detection.Proposed by Alzantot et.al \cite{alzantot2012uptime}.}
\end{figure}

For the step detection in this experiment, acceleration along the walking direction was selected as the signal for thresholding instead of the vertical component. This selection was done after the observation of both components, where the vertical displacement acceleration component showed more variations and noise due to shake, which made it difficult to be used in the detection using FSM approach.\\
To calculate the step length, we used equation (4), which was proposed by Scarlet \cite{scarlett2007enhancing}. This equation uses average, maximum and minimum acceleration values of a step which are denoted by $A_{avg}$, $A_{max}$ and $A_{min}$ respectively. The constant $K$ must be estimated experimentally. \\
\begin{equation}
L = K \frac{(A_{avg} - A_{min})}{(A_{max}-A_{min})} 
\end{equation}
\subsection{Experiments}
Experiment were performed in two phases named: parameter optimization phase and test phase. Walking data obtained in the parameter optimization phase was used to find the optimized values for the necessary parameters and then they were used directly in the next phase to calculate and compare the results. \\
In the parameter optimization phase, actual measurements using Hagisonic device and smart phone measurements for 15 walks were obtained. These data were closely analyzed to find the best possible experimental values for the following variables.
\begin{itemize}
\item{Four thresholds for the FSM step detection}
\item{Parameters of the velocity model}
\item{Constant K of the Scarlet method in step length estimation}
\end{itemize}
These optimized values were fixed for use in the next phase which consisted of walks performed for the comparison. Variation of accuracy, with the amount of corrections in the proposed method was investigated. For this, `Correction Percentage' was defined as the percentage of sensor reading samples per step at which corrections were applied and they were equally spaced for that step.  Results were then compared against each model and SHS systems and is presented in the next section.

\section{Results}
The initial parameter optimization phase consisted of walks of 15m in length which were used to deduce the necessary parameters. These optimized values were used in the next test phase which consisted of 25 walks of 40m length each. \\
Total accuracy is directly affected by the number of steps in the walk for both INS and SHS techniques. Hence the accurate detection of steps is important in these applications. Acceleration signals in the first phase were analyzed individually for the determination of threshold values, which were then averaged to reach the final values for the next phase. In this process, the lower positive threshold ( ``Thr'' according to FSM in figure 2) was carefully selected to be a lower value, since this parameter mainly determines the start of a step. Detection and the threshold values are shown in figure 3 for a walk of 20 steps. 

\begin{figure}[!t]
\centering
\includegraphics[width=2.8in]{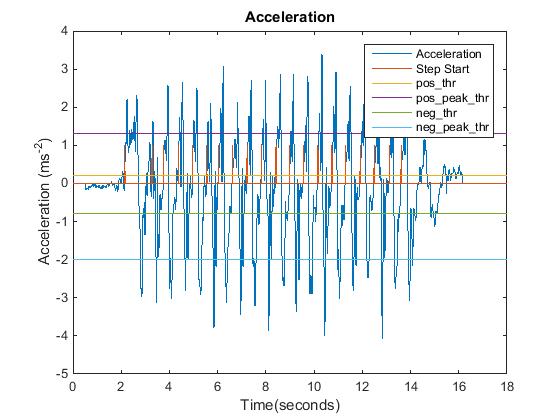}
\caption{Detected steps by FSM algorithm with the acceleration signal and the threshold values for a walk with 20 steps}
\end{figure}

When the deduced thresholds were used in the FSM algorithm, step detection accuracy was 98.6\% in the parameter optimization phase and 98.7\% in the actual test phase. \\
Constant K value in equation (4) for step length estimation was calculated individually for the walks in parameter optimization phase to minimize the distance error. These individual values were then averaged in reaching the fixed constant K for the next phase. \\
The actual velocity measurements were calculated by differentiation using central approximation of the indoor positioning device measurements. Both the matching of the look of actual velocity profile and the minimization of positioning error were considered in selecting the model parameters by using the actual measurements as well as the velocity obtained by integration of the acceleration signal. \\
The velocity shift, mean fraction constant and step fraction constant parameters of Gaussian model were selected as 0.9, 0.4 and 0.15 respectively to match the shape of the model velocity to actual velocity. Then the constant $K$ was obtained by averaging the individual $K$ values which minimized the position error of each walk in parameter optimization phase.
In the Sinusoidal model, amplittude constant was used as 0.25 and shift scale factor was obtained as 1.29 experimentally in parameter optimization phase.
In Sawtooth model, we used the value of 0.2 for amplitude and 0.25 for width constant. Shift scale factor was obtained as 1.67 experimentally in initial phase.\\
Because of the bias and random errors of the sensor units, it is common to have a large amount of drift within about one second. In this experiment, with the basic double integration of acceleration signal, we observed an average error of 34.6m (\(\pm\)4.93m) for a 15m distance walk and an average error of 149.5m (\(\pm\)13.56m) for a 40m distance walk. We propose to minimize this error by introducing an external correction derived from a velocity model. As expected, acceracy improved drastically when the correction was applied. Also the accuracy was high when applied with the complementary Kalman filter than the naive application. The average error was 1.8m (\(\pm\)0.69m) in Kalman filter application for for Gaussian model where it was 2.1m (\(\pm\)0.95m) in naive application with a 10\% correction percentage. \\
The velocity curve for a 15m walk is shown in figure 4. This shows how a large drift occurs in basic integration by the accumulation of small velocity errors which arise due to errors in acceleration readings. Then the velocity is aligned to a correct curve when the corrections are applied. The position is also corrected along with the velocity correction based on the relation defined by Kalman filter transfer function. 

\begin{figure}[!t]
\centering
\includegraphics[width=2.8in]{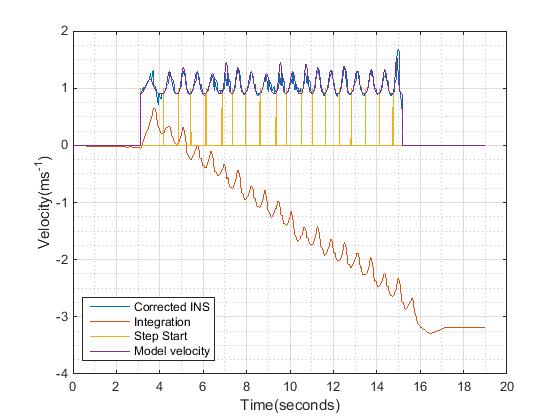}
\caption{Estimated velocity for standard INS vs proposed method with corrections applied using Gaussian velocity model}
\end{figure}

Figure 5 shows the average accuracy and standard deviation values of INS with three different correction models and SHS method with the correction percentage of a step. \\
\begin{figure}[!t]
\centering
\includegraphics[width=2.8in]{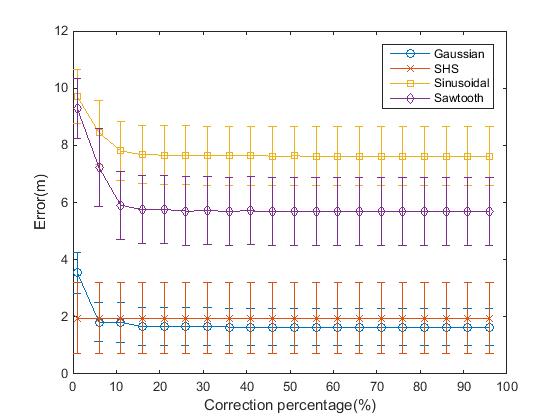}
\caption{Comparison of SHS method and proposed method with three different models showing error vs correction percentage}
\end{figure}
Out of the three models, Gaussian model resulted in highest accuracy compared with others. We hypothesize that this is mainly due to its flexibility in matching the actual velocity curve by using the available parameters. From these results we observe that the proposed method with Gaussian model correction ourperforms the SHS method with increased average accuracy and less standard deviation. The average error with 15\% correction for 40m walk is 1.64.m (\(\pm\)0.66m) in proposed method with Gaussian while it is 1.94m (\(\pm\)1.24m) in SHS method. We can deduce that the proposed method is more robust than SHS as observed by the relatively small standard deviation in the proposed method. The accuracy variation with correction percentage shows that minimum of 5\% correction percentage is sufficient for improved accuracy and no significant improvement after 15\%  \\
With the results from this experiment, we can deduce that INS system on smart phone with corrections improves upon the accuracy and robustness of the SHS method. 

\section{Conclusion}
Recent technological advancements and popularity of smart phones have focussed the research and applications more towards smart phone based indoor localization and tracking. Also with the practical realization advancements of the IOT, knowing the location of devices is important. In this paper, INS approach with correction from a velocity model was suggested and the feasibility of proposed method was investigated experimentally. While the direct double integration does not perform well due to the accumulation of sensor errors, introduction of external corrections in to the basic INS allows it to outperform SHS approach.  Even though the analysis was done offline on a PC, heavy signal processing techniques were avoided in this experiment to make sure that the implementation will fit in to a smart phone platform which are typically constrained by processing and memory resources. In this experiment, better results were achieved by just using a simple velocity model for velocity corrections. Opportunistic use of other external measurements like GPS fixes or even signal fingerprint measurements will allow the proposed algorithm perform with better accuracy for a indoor application. Improvements in modelling of actual velocity and opportunistic use of other external measurements are the next improvements which are on research on this proposed approach for indoor localization and tracking.

\section{Acknowledgement}
This work was supported by NSERC (Natural Sciences and Engineering Research Council of Canada) and University of Western Ontario.





\bibliographystyle{IEEEtran}
\bibliography{references}
%



\end{document}